\documentclass[twocolumn,pre,showpacs]{revtex4}
\usepackage{graphicx}
\usepackage{amsmath}
\usepackage{amsfonts}
\usepackage{mathtools}
\usepackage{subfigure}

\begin{document}
\title{Stochastic thermodynamics of periodically driven systems:\\ 
Fluctuation theorem for currents and unification of two classes}

\author{Somrita Ray and Andre C. Barato}
\affiliation{Max Planck Institute for the Physics of Complex Systems, N\"othnizer Strasse 38, 01187 Dresden,Germany}

\parskip 1mm
\def\F{\mathcal{F}}
\def\g{g_{\mathbf{k},\mathbf{l}}}
\def\gt{\tilde{g}_{\mathbf{k},\mathbf{l}}}

\begin{abstract}
Periodic driving is used to operate machines that go from standard macroscopic engines to small non-equilibrium micro-sized systems. Two classes of such systems are small heat engines driven by 
periodic temperature variations and molecular pumps driven by external stimuli. Well known results that are valid for nonequilibrium steady states of systems 
driven by fixed thermodynamic forces, instead of an external periodic driving, have been generalized to periodically driven heat engines only recently. These results include a general expression for  entropy 
production in terms of currents and affinities and symmetry relations for the Onsager coefficients from linear response theory. For nonequilibrium steady states, the Onsager 
reciprocity relations can be obtained from the more general fluctuation theorem for the currents. We prove a fluctuation  theorem for 
the currents for periodically driven systems. We show that this fluctuation theorem implies a fluctuation dissipation relation, symmetry relations for Onsager coefficients 
and further relations for nonlinear response coefficients. The setup in this paper is more general than previous studies, i.e., our results are valid for both heat engines and molecular pumps. 
The external protocol is assumed to be stochastic in our framework, which leads to a particularly convenient way to treat periodically driven systems.
\end{abstract}
% Explanation of PACS numbers:
% 87.10.Vg: Biological information 
% 05.70.Ln: Nonequilibrium and irreversible thermodynamics
% 02.50.Ey: Stochastic processes 

\maketitle
%==================================================================================================================================================
\section{Introduction}
%==================================================================================================================================================

Thermodynamic cycles of macroscopic systems directed by periodic variation of parameters such as pressure, temperature, and volume, were a primary 
motivation for the  development of the classical theory of thermodynamics \cite{call85}.  The generalization of thermodynamics to 
systems that can have large fluctuations and can be arbitrarily far from equilibrium is a current active area of research known as stochastic 
thermodynamics \cite{seif12}. This theoretical framework is equipped with the tools to deal with periodically driven systems that are  small, 
are far from equilibrium, and operate under finite time conditions. Two main classes of such systems that have been realized experimentally 
are heat engines that are driven by a periodic temperature variation \cite{stee11,blic12,mart15,mart16,mart17} and artificial molecular pumps that generate 
internal net motion due to periodic modulation of energies and energy barriers \cite{leig03,eelk06,li15,erba15}.

The expression of the entropy production in terms of currents (or fluxes)
and affinities \cite{schn76}, and the reciprocity relation of Onsager coefficients \cite{onsa31,onsa31a} are two known 
fundamental results valid for nonequilibrium steady states, which in contrast to periodically driven 
systems are driven by fixed thermodynamic forces. This second result is a cornerstone of  
linear irreversible thermodynamics \cite{groo13}, an older framework that applies to nonequilibrium systems in the linear response regime. 

As an important theoretical advancement for periodically driven heat engines, a general expression of the entropy production in 
terms of currents (or fluxes) and affinities and symmetry relations for the Onsager coefficients have been recently obtained 
in \cite{bran15}. Further general results concerning the linear response regime of periodically driven systems have been derived in 
\cite{proe15,proe16}. Periodically driven heat engines have also been analyzed in several models in the linear response regime \cite{izum09,izum10,izum15}
and arbitrarily far from equilibrium \cite{schm08,espo10,tu14}.

For periodically driven molecular pumps, if the system has an internal fixed load, the periodic driving can lead to output work
against this load. A key difference between this situation and the theoretical approaches considered in \cite{bran15,proe15,proe16}
is that in this case there is a fixed thermodynamic force, i.e., the system would be out of equilibrium even with no periodic 
variation of parameters. Such molecular pumps (also known as ``stochastic pumps'' \cite{astu08}) have received much attention in
recent theoretical studies \cite{raha08,cher08,maes10,raha11,mand14,verl14,asba15,espo15,raz16,bara16a,rost17}.

The fluctuation theorem for the currents is a central result in stochastic thermodynamics valid for nonequilibrium steady states \cite{lebo99,andr07c}
(see \cite{pole14} for a finite time generalization). This result can be expressed as a symmetry on the scaled cumulant generating function of the currents. It 
implies the Onsager reciprocity relations and further relations for nonlinear response coefficients \cite{andr07b}. In this paper, 
we prove a fluctuation theorem for the currents for periodically driven systems. We show that this fluctuation theorem 
implies a fluctuation dissipation relation for periodically driven systems, a symmetry of the Onsager coefficients 
and further relations for nonlinear response coefficients. Our result on the symmetry of Onsager coefficients is a generalization 
of the symmetry from \cite{bran15} for heat engines to a case that also includes molecular pumps.

In our approach we consider discrete state Markov processes with a stochastic protocol \cite{verl14,bara16a}, instead of the more usual 
deterministic protocol. Systems driven by such stochastic protocols have been realized experimentally \cite{gome10,diet15}. The use of a stochastic 
protocol is a mathematical convenience, since in this case the protocol and system together form a bipartite 
Markov process \cite{hart14,bara14a,horo14}. The periodically driven system is then analyzed within the steady state of this bipartite Markov process.
We provide evidence that our results are also valid for deterministic protocols,  which are modeled as a stochastic protocol
with a large number of jumps. We note that a fluctuation theorem for currents for periodically driven systems with a deterministic protocol 
has been proven in \cite{sini11}. Their result is more restrictive than ours as it requires the transition rates to fulfill some constraints 
that, for example, do not allow for the realization of a molecular pump that generates an internal current.  
  
The structure of the paper is as follows. In Sec. \ref{sec2} we define the basic setup and write down an expression for the 
entropy production in terms of currents and affinities. The fluctuation theorem for the currents is proved in Sec. \ref{sec3}. 
The response relations, including the symmetry of the Onsager coefficients are derived in Sec. \ref{sec4}.
We conclude in Sec. \ref{sec5}. The limit of a deterministic protocol is discussed in App. \ref{appa}. Technical aspects of the proof 
of the fluctuation theorem for the currents are discussed in App. \ref{appb}.

%==================================================================================================================================================
\section{General Framework}
%==================================================================================================================================================
\label{sec2}
%==================================================================================================================================================
\subsection{Transition rates and generalized detailed balance}
%==================================================================================================================================================

\begin{figure}
\includegraphics[width=75mm]{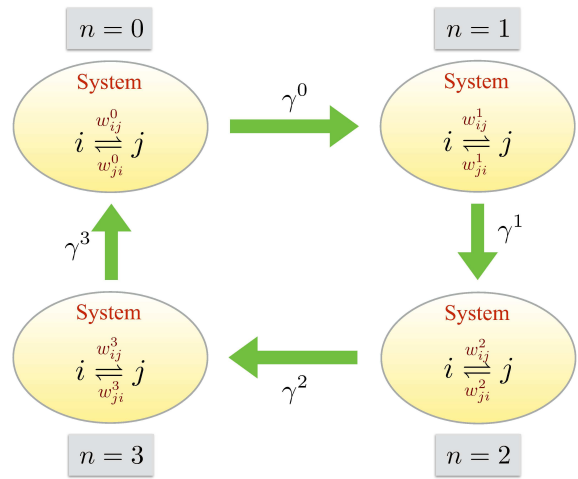}
\vspace{-2mm}
\caption{(Color online) Periodically driven system with a stochastic protocol modelled as a
bipartite Markov process. For this case the number of different states of the external 
protocol is $N=4$. Transition rates that change the state of the system $w_{ij}^n$ depend of the state of the 
external protocol, whereas a transition rate that changes the state of the protocol $\gamma^n$ is independent of the 
state of the system.}
\label{fig1} 
\end{figure}

The system and protocol together form a bipartite Markov process,
which can be used to analyze thermodynamic systems driven by a stochastic protocol \cite{verl14,bara16a}. The variables $i,j$
represent a state of the system, which has a finite number of states $\Omega$. The variable $n=0,1,\ldots,N-1$ represents a 
state of the periodic protocol, as shown in Fig. \ref{fig1}. This variable $n$ is analogous to the time in a periodically 
driven system with a deterministic protocol leading to time-dependent transition rates. 

The transition rate from state $(i,n)$ to state $(j,n)$ is denoted $w_{ij}^n$. If  $w_{ij}^n\neq 0$
then $w_{ji}^n\neq 0$. The transition rate for the protocol in state $n$ to the protocol in state $n+1$ with the system in state $i$
is $w_{i}^{nn+1}= \gamma^n$, while the reversed transition rate is zero. This transition rate is independent of the state of the system $i$,
and from $n=N-1$ the protocol transitions back to state $n=0$. All other rates for transitions that involve a change in the protocol are zero.
The stationary master equation for the whole bipartite process of system and protocol together reads 
\begin{equation}
\frac{d}{dt}P_i^n=\sum_{j}\left(P_j^n w_{ji}^n-P_i^n w_{ij}^n\right)+\gamma^{n-1}P_i^{n-1}-\gamma^nP_i^n=0,
\label{eqME}
\end{equation}
where $P_i^n$ is the stationary probability of state $(i,n)$.

Thermodynamic quantities such as temperature and energy are defined in the following way.  
The energy of state $i$ with the protocol in state  $n$ is
\begin{equation}
E_i^n= E_i+ \Delta E  f_i^n.
\label{eqE}
\end{equation}
The dimensionless function $f_i^n$ characterizes the influence of the external protocol on 
the energy. The energy  $\Delta E$ quantifies the amplitude of the part of the energy that depends 
on the external protocol. The periodicity of the external protocol, as depicted in Fig. \ref{fig1}, 
implies $f_i^{n+N}=f_i^n$. The inverse temperature $\beta^n$ can take values between a hot inverse 
temperature $\beta_h$ and cold inverse temperature $\beta_c\ge \beta_h$. It is written as 
\begin{equation}
\beta^n= \beta_c(1-\F_q h^n),
\label{eqbeta}
\end{equation}
where $h^n\le 1$ and $\F_q\equiv (\beta_c-\beta_h)/\beta_c$. The periodic function $h^{n+N}=h^n$ characterizes the dependence of the temperature on the external protocol. 
Similar forms for the dependence of  energy and temperature on the external protocol for the case of a deterministic protocol have been used in \cite{bran15,proe16}. 
The comparison between a stochastic protocol and a deterministic protocol is discussed in App. \ref{appa}.

The transition rates for changes in the state of the system fulfill the generalized detailed balance relation \cite{seif12}
\begin{equation}
\ln\frac{w_{ij}^n}{w_{ji}^n}= \beta^n\left[E_i^n-E_j^n+(\beta_c)^{-1}\sum_\alpha\F_\alpha d_{ij}^{(\alpha)}\right],
\label{eqgendb}
\end{equation}
where $\F_\alpha$ are internal affinities and $d_{ij}^{(\alpha)}=-d_{ji}^{(\alpha)}$ are 
generalized dimensionless distances. For example, if $\F_\alpha$ is a torque applied to 
a rotatory motor then $d_{ij}^{(\alpha)}$ is the amount that the angle changes in a transition  
from $i$ to $j$. For a heat engine all $\F_\alpha$ are zero. A molecular pump corresponds to the 
case of a fixed temperature $\beta^n=\beta_c$ and non-zero internal force $\F_\alpha$. The comparison 
between Eq. \eqref{eqgendb} and the standard form of the generalized  detailed balance relation for a deterministic protocol 
is presented in App. \ref{appa}.

%==================================================================================================================================================
\subsection{Currents and affinities}
%==================================================================================================================================================

The mathematical form of the rate of entropy production, i.e., the rate of entropy increase of the external medium, reads \cite{bara16a} 
\begin{equation}
\sigma\equiv \sum_n\sum_{ij} P_i^nw_{ij}^n\ln\frac{w_{ij}^n}{w_{ji}^n}\ge 0.
\label{eqent}
\end{equation}
The class of Markov processes considered here is different from the class of Markov processes 
considered in standard stochastic thermodynamics \cite{seif12}. In particular,
transitions that change the state of the external  protocol are irreversible and their transition rates do not appear in Eq. \eqref{eqent}. 
We note that, as usual in thermodynamics, the thermodynamic cost of the external protocol is not taken into account in this paper. Hence, 
the second law in Eq. \eqref{eqent} applies to a non-autonomous physical system, like a heat engine driven by an external control of the 
temperature. The cost of the external protocol becomes relevant if the external control is exerted by, for example, a chemical reaction.
In this case, one must consider a thermodynamically consistent external protocol without irreversible jumps, which leads to a different 
statement of the second law \cite{bara17}.

The average elementary probability current from state $(i,n)$ to state $(j,n)$ is defined as 
\begin{equation}
J_{ij}^n\equiv P_i^n w_{ij}^n-P_j^n w_{ji}^n.
\label{eqelem}
\end{equation}
The rate of entropy production in Eq. \eqref{eqent} in terms of this elementary probability current
becomes
\begin{equation}
\sigma=\sum_n\sum_{i<j} J_{ij}^n\ln\frac{w_{ij}^n}{w_{ji}^n},
\end{equation}
where the sum $\sum_{i<j}$ is over all links between states of the system. 
Using the generalized detailed balance relation in Eq. \eqref{eqgendb} we obtain
\begin{equation}
\sigma=\sum_\alpha \F_\alpha J_\alpha + \sum_n\sum_{i<j} J_{ij}^n\beta^n(E^n_i-E^n_j),
\label{eqent2} 
\end{equation}
where
\begin{equation}
J_\alpha\equiv \sum_{n}\sum_{i<j} (\beta_c)^{-1}\beta^nJ_{ij}^nd_{ij}^{(\alpha)}. 
\label{eqjalpha}
\end{equation}

Using Eq. \eqref{eqbeta}, the second term on the right hand side of Eq. \eqref{eqent2}  
becomes
\begin{equation}
\sum_n\sum_{i<j} J_{ij}^n\beta^n (E^n_i-E^n_j)= \beta_c\sum_n\sum_{i<j} J_{ij}^n (E^n_i-E^n_j)+ \F_q J_q,
\label{eqsecondterm}
\end{equation}
where 
\begin{equation}
J_q\equiv \sum_{n}\sum_{i<j} J_{ij}^n h^n\beta_c(E_j^n-E_i^n).
\label{eqjq} 
\end{equation}
This current is the generalized heat flux from \cite{bran15}. For the case of $\F_\alpha=0$ and a temperature that takes only 
the values $\beta_c$ (for $h^n=0$) and $\beta_h$ (for $h^n=1$), $J_q$ is the rate at which heat is taken from 
the hot reservoir multiplied by $\beta_c$. 

The work current $J_e$ is defined as 
\begin{align}
J_e &\equiv  \sum_n\sum_{i<j} J_{ij}^n (f^n_i-f^n_j)\nonumber\\
& =\sum_n\sum_{i} P_i^n \gamma^n(f^{n+1}_i-f^n_{i}), 
\label{eqje} 
\end{align}
where the second equality follows from the master equation in Eq. \eqref{eqME}, which leads to 
$\frac{d}{dt}\sum_{i}\sum_{n}f_i^nP_i^n=0$. The term $\Delta E J_e$ is the rate of work exerted on 
the  system due to the variation of the external protocol: from the second line of Eq. \eqref{eqje}, 
$\gamma^n$ is the speed of the change of the protocol from $n$ to $n+1$ and $\Delta E(f^{n+1}_i-f^n_{i})$
is the energy change associated with the protocol jump. Finally, using Eqs. \eqref{eqent2}, 
\eqref{eqsecondterm}, \eqref{eqje}, and the dimensionless affinity
$\F_e= \beta_c\Delta E$ we obtain   
\begin{equation}
\sigma= \F_q J_q+ \F_eJ_e+\sum_\alpha \F_\alpha J_\alpha,
\label{entprodfin}
\end{equation}
which is the expression of the entropy production in terms of currents and affinities.  Note that 
we have defined the currents in Eqs. \eqref{eqjalpha}, \eqref{eqjq}, and \eqref{eqje}, in such a way that the 
the affinities $\F_\alpha$, $\F_q$, and $\F_e$ are dimensionless. The comparison between this expression for $\sigma$
and the more usual expression for the entropy production for a deterministic protocol is discussed in App. \ref{appb}.  
In order to illustrate the general theory we introduce two specific models: one for a heat engine and one for molecular pump.

%==================================================================================================================================================
\subsection{Illustrative examples}
%==================================================================================================================================================

\subsubsection{Heat Engine}

\begin{figure}
\includegraphics[width=75mm]{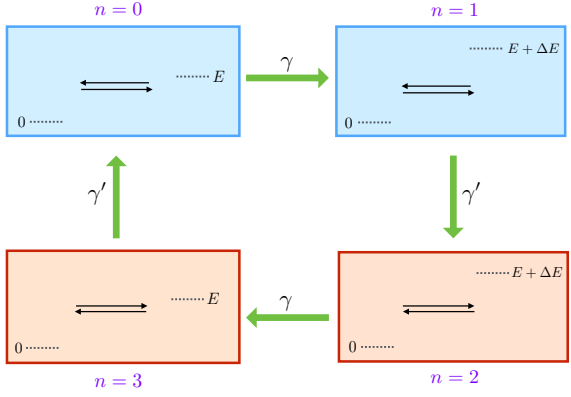}
\vspace{-2mm}
\caption{(Color online) Model for a heat engine. The temperature is cold for $n=0,1$ and hot for $n=2,3$.
The transition rate from the state with energy $0$ to the 
states with energy $E^n$ is set to $k\textrm{e}^{-\beta^nE^n/2}$, while the reversed transition rate is 
$k\textrm{e}^{\beta^nE^n/2}$. The transition rate associated with isothermal changes is $\gamma$, whereas the
transition rate associated with temperature changes is $\gamma'$.
}
\label{fig2new} 
\end{figure}

The model for a heat engine is illustrated in Fig. \ref{fig2new}. The system has two states, a down state with energy 0 and an up state with energy $E^n=E+\Delta Ef^n$. The 
protocol has four states. The first jump of the protocol corresponds to an isothermal step at temperature $\beta_c$, with the energy of the up state lifted from $E$
to $E+\Delta E$. In the second jump of the protocol the temperature is changed from $\beta_c$ to $\beta_h$. In the third jump, the energy is lowered 
back from $E+\Delta E$ to $E$ in an isothermal process at temperature $\beta_h$. In the fourth jump, the engine returns to the initial state, with a temperature change from 
$\beta_h$ to $\beta_c$. In the isothermal steps, work is exerted on the system when the higher energy level is elevated by $\Delta E$ at temperature 
$\beta_c$ and work is extracted from the system when the higher energy level is lowered at temperature $\beta_h$. If the temperature difference is high 
enough, the system is more likely to be in the state of higher energy during the work extraction step, leading to net work extraction. For this model,
$f^n=\delta_{n,1}+\delta_{n,2}$ and $h^n$ from Eq. \eqref{eqbeta} is $h^n=\delta_{n,2}+\delta_{n,3}$.

The entropy production for the heat engine reads
\begin{equation}  
\sigma= \F_q J_q+ \F_eJ_e,
\end{equation}
where $J_q$ is the rate of heat taken from the hot reservoir and $-\F_e J_e$ is the rate of extracted work, both in units of $\beta_c^{-1}$ per time. 
Taking the transition rates given in the caption of Fig. \ref{fig2new} we consider the following limit. First, we take the limit at which temperature changes 
are instantaneous, leading to $\gamma'\gg \gamma,k$. Second, we consider that the system equilibrates before an isothermal step, i.e., $k\gg\gamma$.
Within this limit, calculating the stationary distribution of the full bipartite system  we obtain the following simple expressions 
\begin{equation}  
-J_e= \gamma \frac{\textrm{e}^{\beta_cE}-\textrm{e}^{\beta_h(E+ \Delta E)}}{2(1+\textrm{e}^{\beta_h(E+ \Delta E)})(1+\textrm{e}^{\beta_cE})}
\end{equation}  
and 
\begin{equation}  
J_q= \gamma\beta_c(E+\Delta E)  \frac{\textrm{e}^{\beta_cE}-\textrm{e}^{\beta_h(E+ \Delta E)}}{2(1+\textrm{e}^{\beta_h(E+ \Delta E)})(1+\textrm{e}^{\beta_cE})},
\end{equation}  
which leads to the entropy production 
\begin{equation}  
\sigma= \gamma[\beta_cE-\beta_h(E+ \Delta E)]\frac{\textrm{e}^{\beta_cE}-\textrm{e}^{\beta_h(E+ \Delta E)}}{2(1+\textrm{e}^{\beta_h(E+ \Delta E)})(1+\textrm{e}^{\beta_cE})}\ge 0.
\label{eqsecheat}
\end{equation}  
Hence, for $\beta_h/\beta_c\le E/(E+ \Delta E)$ this machine operates as a heat engine that uses part of the heat taken from the hot reservoir to extract work. 
Interestingly, the efficiency of the heat engine in this regime is independent of the temperature difference, i.e., 
\begin{equation}  
\eta\equiv \frac{-\F_eJ_e}{J_q}= \frac{\Delta E}{E+\Delta E}\le 1-\frac{\beta_h}{\beta_c}.
\end{equation} 
The second inequality, which follows from the second law in Eq. \eqref{eqsecheat}, tells us that the efficiency 
of the heat engine is bounded by the Carnot efficiency.     
    
\subsubsection{Molecular pump}

\begin{figure}
\includegraphics[width=75mm]{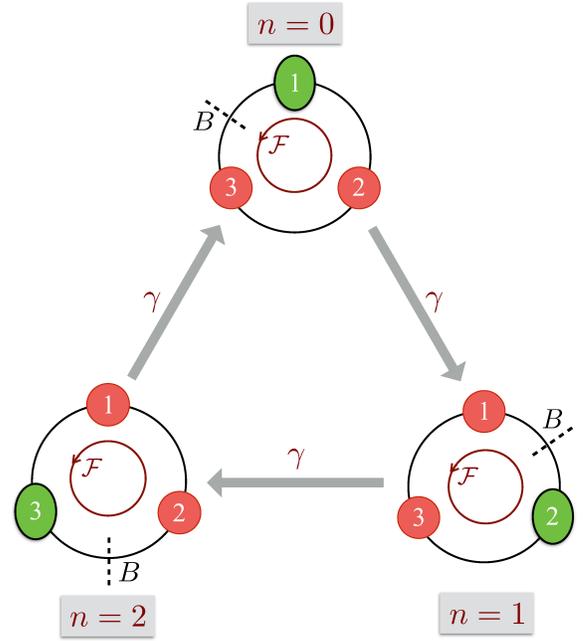}
\vspace{-2mm}
\caption{(Color online) Model for a molecular pump. The ellipse in green represents a state with energy $\F_e$ and the red circles 
represent states with energy $0$. The transition rates for $n=0$ are $w_{12}^0= k \textrm{e}^{\F_e-\F/3}$,  
 $w_{13}^0= k \textrm{e}^{\F_e-B}$,   $w_{23}^0= k\textrm{e}^{-\F/3}$, $w_{21}^0= k$, $w_{31}^0= k\textrm{e}^{-B-\F/3}$, and $w_{32}^0= k$,
where $\beta_c=1$. Changing $n$ leads to a rotation in the clock wise direction of the transition rates. For example, $w_{12}^0=w_{23}^0=w_{31}^0$.}
\label{fig4} 
\end{figure}

We consider a model for a molecular pump shown in Fig. \ref{fig4}, which has a protocol with $N=3$ states and 
$\Omega=3$ internal states, where $i=1,2,3$. This model has been analyzed in \cite{bara16a,bara17}. The temperature is fixed 
and set to $\beta^n=1$.  The energy is set to  $E_i^n=\F_e\delta_{i,n+1}$,
i.e., the green state in Fig. \ref{fig4} has energy   $\F_e$ and the other two states have energy $0$. The dotted line in Fig. \ref{fig4}
represents an energy barrier $B$. The transition rates for a change 
in the external protocol are all $\gamma^n=\gamma$. The internal transition rates fulfilling the generalized detailed balance relation in Eq. \eqref{eqgendb} are given in
the caption of Fig. \ref{fig4}. 

The clockwise rotation of both this energy barrier and the state with higher energy can lead to an internal current in the clockwise direction that goes against an 
internal load $\F$ in the anticlockwise direction. For such a molecular pump the entropy production in Eq. \eqref{entprodfin} takes the form 
\begin{equation}
\sigma= J_e\F_e+J_\alpha\F
\end{equation}
where $J_\alpha$ is the internal current defined in Eq. \eqref{eqjalpha}, with $d_{ij}^\alpha=1/3$ for a clockwise transition and $d_{ij}^\alpha=-1/3$ for a anti-clockwise transition.
The work exerted on the system $J_e\F_e$ can lead to work done against the internal force $-J_\alpha\F$, with an efficiency $\eta\equiv (-J_\alpha\F)/(J_e\F_e)$. 
In the limit of an infinite energy barrier $B$ and for internal transitions that are much faster than changes in the external protocol ($k>>\gamma$), we obtain 
the following expressions for the currents    
\begin{equation}  
-J_\alpha= \gamma\frac{\textrm{e}^{\F/3+\F_e}+\textrm{e}^{\F_e}-2\textrm{e}^{2\F/3}}{3(\textrm{e}^{\F/3+\F_e}+\textrm{e}^{\F_e}+\textrm{e}^{2\F/3}) }
\end{equation}
and  
\begin{equation}  
J_e= \gamma\frac{\textrm{e}^{\F/3}\left(\textrm{e}^{\F_e}-\textrm{e}^{\F/3}\right)}{3(\textrm{e}^{\F/3+\F_e}+\textrm{e}^{\F_e}+\textrm{e}^{2\F/3}) }.
\end{equation}  
Therefore, for a fixed positive $\F_e$, this model operates as a molecular pump that does work against the internal force $0\le\F\le\F^*$, where 
$\F^*$ is the solution of the equation $J_\alpha=0$.

%==================================================================================================================================================
\subsection{Reversed protocol}
%==================================================================================================================================================

\begin{figure}
\includegraphics[width=75mm]{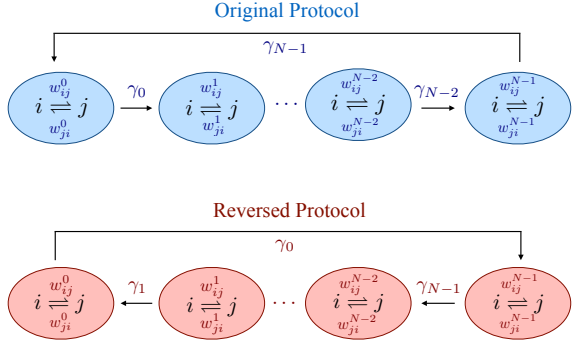}
\vspace{-2mm}
\caption{(Color online) Illustration of the comparison between the original bipartite Markov process with rates in Eq. \eqref{defrates1} and 
the one corresponding to reversal of the protocol with rates in Eq. \eqref{defrates2}.}
\label{fig2} 
\end{figure}

Our results in the next section are obtained in terms of the original bipartite Markov process and another bipartite Markov process that corresponds to 
reversal of the external protocol, which is represented in Fig. \ref{fig2}. The transition rates for the original bipartite process are given by   
 \begin{equation}	
w_{ij}^{nn'}\equiv\left\{
\begin{array}{ll} 
 w^{n}_{ij} & \quad  \textrm{if $i\neq j$ and $n'=n$},\\
 \gamma^n & \quad \textrm{if $i=j$ and $n'=n+1$}, \\
 0 & \quad \textrm{otherwise}. 
\end{array}\right.\,
\label{defrates1}
\end{equation}
The transition rates for the bipartite Markov process that corresponds to reversal of the protocol are
\begin{equation}	
v_{ij}^{nn'}\equiv\left\{
\begin{array}{ll} 
 w^{n}_{ij} & \quad  \textrm{if $i\neq j$ and $n'=n$},\\
 \gamma^n & \quad \textrm{if $i=j$ and $n'=n-1$}, \\
 0 & \quad \textrm{otherwise}. 
\end{array}\right.\,
\label{defrates2}
\end{equation}
For a symmetric protocol, the bipartite Markov processes defined in Eqs. \eqref{defrates1} and \eqref{defrates2}
become equivalent. Such a symmetric protocol fulfills the conditions $w_{ij}^n=w_{ij}^{N-1-n}$ and $\gamma^n=\gamma^{N-1-n}$.

%==================================================================================================================================================
\section{Fluctuation theorem for currents}
%==================================================================================================================================================
\label{sec3}

\subsection{Fluctuating currents}

A fluctuating elementary current $X_{ij}^n$ is a functional of the stochastic trajectory from time $0$ to time $t$ that counts transitions between states $(i,n)$ and $(j,n)$. 
For compact notation, we omit the dependence of $X_{ij}^n$ on the time interval $t$. If a transition from $(i,n)$ to $(j,n)$ happens, this random variable increases by one and 
if a transition from $(j,n)$ to $(i,n)$ happens this random variable decreases by one. The average of this fluctuating current is 
\begin{equation}
\lim_{t\to \infty} \frac{\langle X_{ij}^n\rangle}{t}=J_{ij}^n, 
\end{equation}
where the angular brackets indicate an average over stochastic trajectories. Similar to Eq. \eqref{eqjalpha}, the fluctuating currents $X_\alpha$ are given by
\begin{equation}
X_\alpha\equiv \sum_{n}\sum_{i<j} (\beta_c)^{-1} \beta^nX_{ij}^nd_{ij}^{(\alpha)}. 
\end{equation}
Furthermore, from Eq. \eqref{eqjq} we define
\begin{equation}
X_q\equiv \sum_{n}\sum_{i<j} X_{ij}^n h^n\beta_c(E_j^n-E_i^n) ,
\end{equation}
and from Eq. \eqref{eqje} we define
\begin{equation}
X_e\equiv \sum_{n}\sum_{i<j} X_{ij}^n(f_i^n-f_j^n). 
\end{equation}
The fluctuating entropy production $X_s$ reads 
\begin{equation}
X_s\equiv \F_qX_q+\F_eX_e+\sum_\alpha \F_\alpha X_\alpha=\sum_a\F_aX_a,  
\end{equation}
where the sum $\sum_a$ represents a sum over all currents and affinities including $a=q$, $a=e$, and $a=\alpha$.

The scaled cumulant generating function associated with the vector of currents $\mathbf{X}=(X_a)$ is defined as  
\begin{equation}
G(\mathbf{z})\equiv \lim_{t\to\infty}\frac{1}{t}\ln\langle\exp(\mathbf{z}\cdot\mathbf{X})\rangle, 
\end{equation}
where $\mathbf{z}=(z_a)$ is a vector of real numbers $\mathbf{z}\cdot\mathbf{X}\equiv\sum_a z_aX_a$. This quantity is related to 
the rate function $I(\mathbf{x})$ from large deviation theory \cite{touc09} , which is defined as
\begin{equation}
\textrm{Prob}(\mathbf{X})\sim \exp[-tI(\mathbf{x})], 
\end{equation}
where $\mathbf{x}\equiv \mathbf{X}/t$ and the symbol $\sim$ indicates asymptotic behavior in the limit $t\to \infty$. Specifically, 
$I(\mathbf{x})$ is a Legendre-Fenchel transform of $G(\mathbf{z})$, i.e., 
 \begin{equation}
 I(\mathbf{x}) = \textrm{max}_{\mathbf{z}}\left[\mathbf{x}\cdot\mathbf{z} -G(\mathbf{z})\right].
\label{eqtrans}
\end{equation}

\subsection{Fluctuation theorem}

We now prove the fluctuation theorem for the currents, which is a symmetry in 
the scaled cumulant generating function $G(\mathbf{z})$. The modified generator $\mathcal{L}(\mathbf{z})$
is a quadratic matrix with dimension $\Omega\times N$. Its elements are identified by a state of the bipartite 
process $i,n$. These elements are defined as  
\begin{equation}	
[\mathcal{L}(\mathbf{z})]_{j,n';i,n}\equiv\left\{
\begin{array}{ll} 
 w^{n}_{ij} \textrm{e}^{\sum_ad^{n(a)}_{ij} z_a} &   \textrm{ if $j\neq i$ and $n'=n$},\\
 \gamma^n &  \textrm{ if $j=i$ and $n'=n+1$}, \\
 -\gamma^n-\sum_{k}w_{ik}^n & \textrm{ if $j=i$ and $n'=n$}, \\
 0 &  \textrm{ otherwise}. 
\end{array}\right.\,
\label{eqgen}
\end{equation}
where $d^{n(\alpha)}_{ij}\equiv (\beta_c)^{-1}\beta^nd_{ij}^{(\alpha)}$, $d^{n(q)}_{ij}\equiv h^n\beta_c(E_j^n-E_i^n)$, and $d^{n(e)}_{ij}\equiv (f_i^n-f_j^n)$.
This matrix can be written in the form 
\begin{equation}
\mathcal{L}(\mathbf{z})=\left(\begin{array}{ccccc}
\mathcal{\mathcal{L}}_{0}(\mathbf{z})-\mathbf{\Gamma}_0 & 0 & \ldots & \mathbf{\Gamma}_{N-1}\\
\mathbf{\Gamma}_0 & \mathcal{\mathcal{L}}_{1}(\mathbf{z})-\mathbf{\Gamma}_1 &  \ldots  & 0\\
0 & \mathbf{\Gamma}_1 &  \ldots  & 0\\
\vdots & \vdots & \ddots & \vdots \\
0 & 0 & \ldots &  \mathcal{\mathcal{L}}_{N-1}(\mathbf{z})-\mathbf{\Gamma}_{N-1}
\end{array}\right).
\label{eqgensto1}
\end{equation}
where 
\begin{equation}	
[\mathcal{L}_n(\mathbf{z})]_{j;i}\equiv\left\{
\begin{array}{ll} 
w^n_{ij} \textrm{e}^{\sum_ad^{n(a)}_{ij} z_a}  &  \textrm{if $i\neq j$},\\
 -\sum_{k}w^n_{ik} &  \textrm{ if $i=j$}, 
\end{array}\right.\,
\label{eqgensto2}
\end{equation}
and $\mathbf{\Gamma}_n=\gamma^n\mathbf{I}$, with $\mathbf{I}$ as the identity matrix with dimension $\Omega$.
This modified generator is a Perron-Frobenius matrix, and its maximum eigenvalue is 
the scaled cumulant generating function $G(\mathbf{z})$ \cite{lebo99}.

The scaled cumulant generating function associated with the reversed bipartite process, with transition rates given 
by Eq. \eqref{defrates2}, is denoted $G^R(\mathbf{z})$. The modified generator related to it is 
\begin{equation}	
[\mathcal{L}^R(\mathbf{z})]_{j,n';i,n}\equiv\left\{
\begin{array}{ll} 
 w^{n}_{ij} \textrm{e}^{\sum_ad^{n(a)}_{ij} z_a} &  \textrm{ if $j\neq i$ and $n'=n$},\\
 \gamma^n &  \textrm{ if $j=i$ and $n'=n-1$}, \\
 -\gamma^n-\sum_{k}w_{ik}^n &  \textrm{ if $j=i$ and $n'=n$}, \\
 0 &  \textrm{ otherwise}. 
\end{array}\right.\,
\label{eqgen2}
\end{equation}
This matrix can be written in the form 
\begin{equation}
\mathcal{L}^R(\mathbf{z})=\left(\begin{array}{ccccc}
\mathcal{L}_{0}(\mathbf{z})-\mathbf{\Gamma}_0 & \mathbf{\Gamma}_1 & \ldots & 0\\
0 & \mathcal{L}_{1}(\mathbf{z})-\mathbf{\Gamma}_1 &  \ldots  & 0\\
0 & 0 &  \ldots  & 0\\
\vdots & \vdots & \ddots & \vdots \\
\mathbf{\Gamma}_0 & 0 & \ldots &  \mathcal{L}_{N-1}(\mathbf{z})-\mathbf{\Gamma}_{N-1}
\end{array}\right).
\label{eqgensto3}
\end{equation}

From Eqs. \eqref{eqE}, \eqref{eqbeta}, \eqref{eqgendb}, and \eqref{eqgensto2}  we obtain the following symmetry,
\begin{equation}	
[\mathcal{L}_n(\mathbf{z})]_{j;i}=[\mathcal{L}_n(-\mathbf{F}-\mathbf{z})]_{i;j}\textrm{e}^{\beta_c(E_i-E_j)},
\label{eqsmallsym}
\end{equation}	
where $\mathbf{F}= (\F_a)$ and $E_i$ is the part of the energy $E_i^n$ that does not depend on 
the external protocol. For the case $\gamma^n=\gamma$, with a matrix $\mathcal{D}$ that 
is a diagonal matrix with components $[\mathcal{D}]_{i,n;j,n'}=\delta_{nn'}\delta_{ij}\textrm{e}^{\beta_cE_i}$, 
we obtain 
\begin{equation}	
\mathcal{L}(\mathbf{z})=\left(\mathcal{D}\mathcal{L}^R(-\mathbf{F}-\mathbf{z})\mathcal{D}^{-1}\right)^{T},
\label{eqsimi} 
\end{equation}	
where the superscript $T$ denotes transpose. This similarity transformation proves that $\mathcal{L}(\mathbf{z})$ 
and $\mathcal{L}^R(-\mathbf{F}-\mathbf{z})$ have the same characteristic polynomial.
A similar similarity transformation appears in the proof of a transient fluctuation theorem for the currents \cite{pole14}.

For general $\gamma^n$ Eq. \eqref{eqsimi} does not hold, however, as shown in App. \ref{appb},
the characteristic polynomials of $\mathcal{L}(\mathbf{z})$ and $\mathcal{L}^R(-\mathbf{F}-\mathbf{z})$
are the same. Since the scaled cumulant  generating function is the maximum eigenvalue 
of the modified generator, this equality between characteristic polynomials implies the symmetry
\begin{equation}	
G(\mathbf{z})=G^R(-\mathbf{F}-\mathbf{z}).
\label{eqft}
\end{equation}	
This fluctuation theorem for the currents for periodically driven systems is the most general result of this paper.
It is a generalization of the fluctuation theorem for the currents for nonequilibrium steady states \cite{lebo99,andr07b}
to periodically driven systems. For the case of a symmetric protocol this relation becomes $G(\mathbf{z})=G(-\mathbf{F}-\mathbf{z})$, 
which is the exact same form of the fluctuation theorem for the currents for nonequilibrium steady states. 
In spite of this same form and a similar mathematical derivation, the relation $G(\mathbf{z})=G(-\mathbf{F}-\mathbf{z})$ for symmetric protocols 
is a different mathematical result, which applies to a different class of Markov processes, in relation to the fluctuation theorem for the currents
for nonequilibrium steady states. We point out that our results should also be valid for deterministic protocols that are continuous, since there is 
strong evidence that such protocols can be obtained as a limit of a stochastic protocol with infinitely many jumps, as discussed in App. \ref{appa}.

It is worth mentioning that a fluctuation theorem for currents for a system driven by periodic and deterministic protocols has been obtained in
\cite{sini11}. Their derivation, however, relies on assumptions that restrict the time dependence of transitions rates. In particular, they cannot 
have a situation in which both energies and energies barriers are varied in time, which is a necessary condition for a molecular pump to generate an 
internal current \cite{raha08,cher08}. Hence, the fluctuation theorem from  \cite{sini11} cannot be used to derive the response relations from 
Sec. \ref{sec4} that are valid for both heat engines and molecular pumps. 

\begin{figure}
\subfigure[]{\includegraphics[width=75mm]{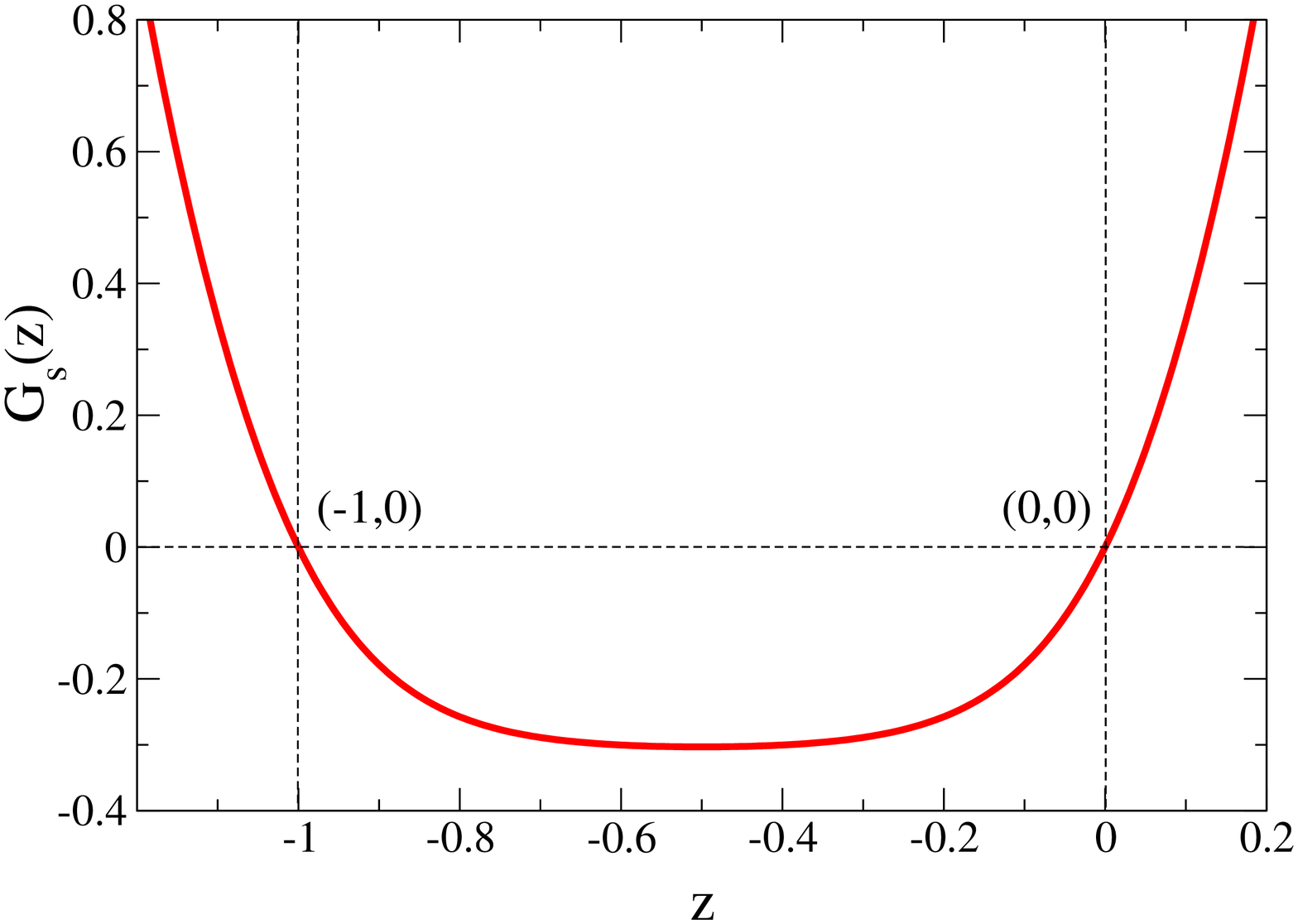}\label{fig5a}}
\subfigure[]{\includegraphics[width=75mm]{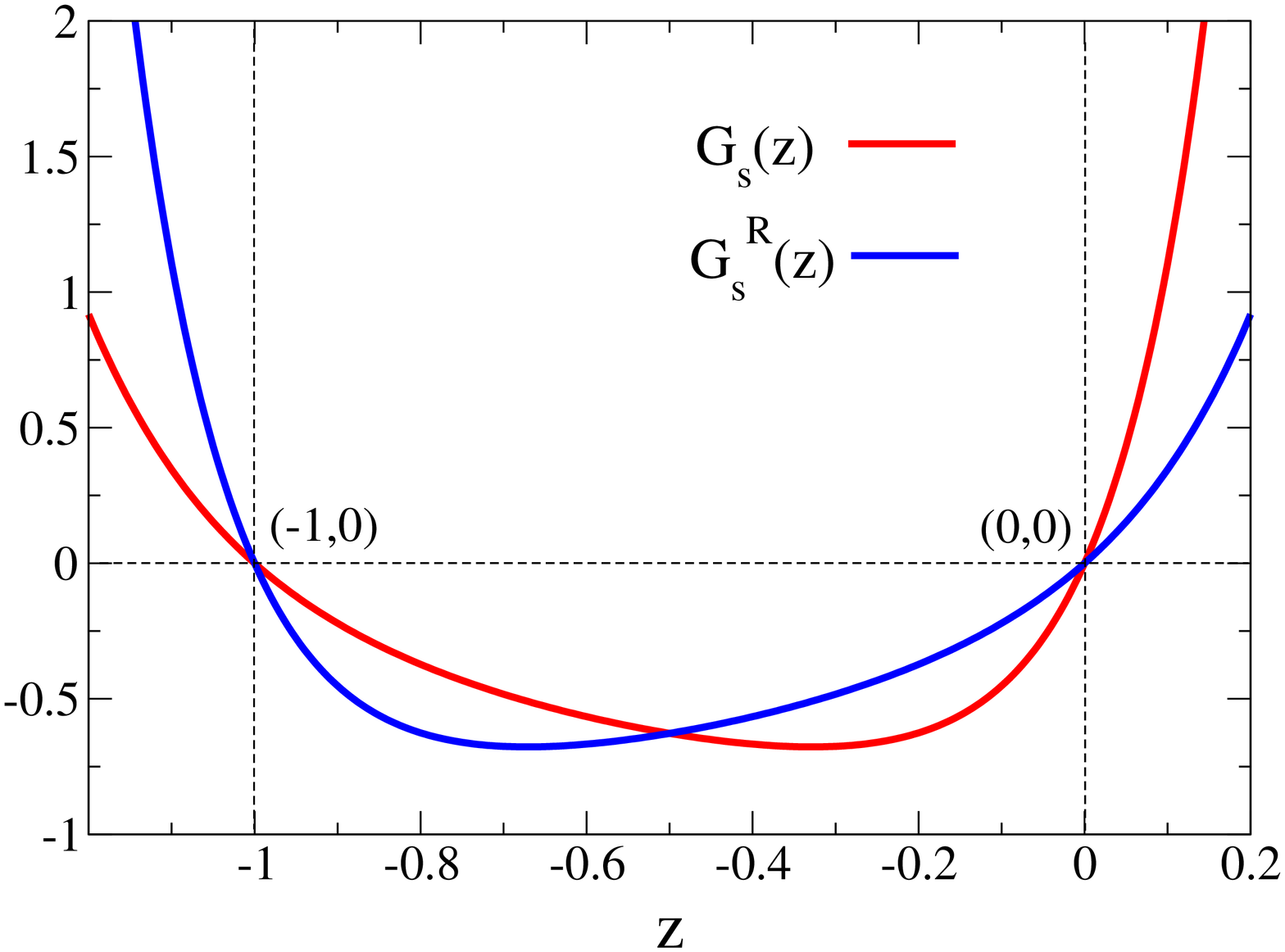}\label{fig5b}}
\vspace{-2mm}
\caption{(Color online) Scaled cumulant generating function associated with the entropy current $G_s(z)$.
(a) Symmetric protocol for the model depicted in Fig. \ref{fig3}. Parameters were set to 
$k=\F_e=10$, $\gamma^0=1$, $\gamma^1=2.5$, $\gamma^2=4$, and $\gamma^3=3$. (b) Non-symmetric protocol for 
the molecular pump depicted in Fig. \ref{fig4}. Parameters were set to $k=10$, $\gamma=1$, $\F_e=B=10$, and $\F=5$.}   
\label{fig5} 
\end{figure}

The scaled cumulant generating function associated with the entropy current $X_s$ is obtained 
by setting the real vector to $\mathbf{z}= (\F_a z)$, i.e., 
\begin{equation}	
G_s(z)=G(z\mathbf{F}).
\label{eqGS}
\end{equation}	
The fluctuation theorem for the currents implies
\begin{equation}	
G_s(z)=G_s^R(-1-z).
\label{eqfts}
\end{equation}	
This equation is a generalization of the Gallavotti-Cohen symmetry \cite{lebo99} to periodically driven systems. 
In Fig. \ref{fig5} we plot $G_s(z)$ for the models explained in App. \ref{appb}. As illustrated in Fig. \ref{fig5a}, the function $G_s(z)$ is symmetric for the case of a symmetric protocol. Furthermore, 
as shown in Fig. \ref{fig5b}, for a non-symmetric protocol $G_s(z)$ fulfills the property $G_s(0)=G_s(-1)=0$, which is a consequence of Eq. \eqref{eqfts}. This property, which is 
also valid for $G(\mathbf{z})$,  is important for the derivations in the next section. We note that in terms of the rate function  $I(\mathbf{x})$, the
fluctuation theorem for the currents in Eq. \eqref{eqft} becomes 
\begin{equation}	
I(\mathbf{x})-I^R(-\mathbf{F}-\mathbf{x})=-\mathbf{F}\cdot\mathbf{x},
\end{equation}	
where we have used Eq. \eqref{eqtrans}.

%==================================================================================================================================================
\section{Response coefficients}
%==================================================================================================================================================
\label{sec4}

\subsection{Fluctuation dissipation relation}

In this section we write the scaled cumulant generating function as $G(\mathbf{z},\mathbf{F})$, keeping the dependence on the affinities explicit. 
An average current $J_a$ can be obtained from $G(\mathbf{z},\mathbf{F})$ with the equation
\begin{equation}
J_a(\mathbf{F})=\left.\frac{\partial G}{\partial z_a}\right|_{\mathbf{z}=0}.
\label{eqcurrG}
\end{equation}
Furthermore, the diffusion coefficient  is defined as 
\begin{equation}
D_{ab}(\mathbf{F})\equiv \frac{\langle (X_a-\langle X_a\rangle)(X_b-\langle X_b\rangle)\rangle}{t}=\left.\frac{\partial^2 G}{\partial z_az_b}\right|_{\mathbf{z}=0}.
\end{equation}

In the linear response regime the current in Eq. \eqref{eqcurrG} becomes 
\begin{equation}
J_a= \sum_bL_{ab}\F_b+\textrm{O}(\F^2),
\end{equation}
where
\begin{equation}
L_{ab}\equiv \left.\frac{\partial^2 G}{\partial z_a\partial\F_b}\right|_{\mathbf{z}=0,\mathbf{F}=0}
\end{equation}
are the Onsager coefficients. We now derive a fluctuation dissipation relation for periodically driven systems 
that relates the response coefficients $L_{ab}$ with fluctuations in equilibrium, as quantified by $D_{ab}^{\textrm{eq}}\equiv D_{ab}(\mathbf{F}=0)$.

The fluctuation theorem for the currents \eqref{eqft} implies the relation  
\begin{equation}	
G(0,\mathbf{F})=G(-\mathbf{F},\mathbf{F})=0.
\label{eq0}
\end{equation}	
A Taylor expansion around $\mathbf{z}=\mathbf{F}=0$ of the scaled cumulant generating function leads to
\begin{equation}
G(\mathbf{z}^*,\mathbf{F}^*)= \sum_{\mathbf{k}\mathbf{l}}\g \prod_a \frac{(z_a^*)^{k_a}(\F_a^*)^{l_a}}{k_a!l_a!}
\label{eqtaylor1}
\end{equation}
where 
\begin{equation}
\g\equiv \left.\frac{\partial^{k+l} G}{\prod_a\partial^{k_a}z_a\partial^{l_a}\F_a}\right|_{\mathbf{z}=0,\mathbf{F}=0},
\end{equation}
$\mathbf{k}\equiv (k_a)$, $\mathbf{l}\equiv (l_a)$, $k=\sum_a k_a$, and $l=\sum_a l_a$. The sum $\sum_{\mathbf{k}\mathbf{l}}$
is over all possible vectors with each component taking the values $k_a=0,1,\ldots,\infty$ and $l_a=0,1,\ldots,\infty$.
With a Taylor expansion around $-\mathbf{z}^*-\mathbf{F}^*$, we obtain
\begin{align}
& G(-\mathbf{F}^*-\mathbf{z}^*,\mathbf{F}^*)= \sum_{\mathbf{k}\mathbf{l}}\g \prod_a \frac{(-z_a^*-\F_a^*)^{k_a}(\F_a^*)^{l_a}}{k_a!l_a!}\nonumber\\
& =\sum_{\mathbf{k}\mathbf{l}}\gt \prod_a \frac{(-z_a^*)^{k_a}(\F_a^*)^{l_a}}{k_a!l_a!},
\label{eqtaylor2}
\end{align}
where
\begin{equation}
\gt\equiv \left.\frac{\partial^{k+l} G}{\prod_a\partial^{k_a}z_a\partial^{l_a}\F_a}\right|_{\mathbf{z}=-\mathbf{F^*},\mathbf{F}=0}.
\end{equation}
Eq. \eqref{eqtaylor2} implies  
\begin{equation}
\gt= \sum_{\mathbf{n}}g_{\mathbf{k}+\mathbf{n},\mathbf{l}-\mathbf{n}}\prod_a(-1)^{n_a} \frac{l_a!}{(l_a-n_a)!n_a!},
\label{eqgtg}
\end{equation}
where $\mathbf{n}\equiv (n_a)$ and $n_a=0,1,\ldots,l_a$ in the sum $\sum_{\mathbf{n}}$. The zeros of $G(\mathbf{z},\mathbf{F})$ in Eq. \eqref{eq0}, combined with Eq. \eqref{eqtaylor1} and \eqref{eqtaylor2} 
lead to $g_{0,\mathbf{l}}=\tilde{g}_{0,\mathbf{l}}=0$ for all vectors $\mathbf{l}$. Hence, setting $\mathbf{k}=0$ in 
Eq. \eqref{eqgtg},  we obtain
\begin{equation}
 g_{\mathbf{l},0}= -\sum_{\mathbf{n}}\nolimits'g_{\mathbf{n},\mathbf{l}-\mathbf{n}}\prod_a(-1)^{n_a} \frac{l_a!}{(l_a-n_a)!n_a!},
 \label{eqresp1}
\end{equation}
where the sum $\sum_{\mathbf{n}}\nolimits'$ is over  all $n_a=0,1,\ldots,l_a$ apart from the term $n_a=l_a$ for all $a$. A similar mathematical derivation of Eq. \eqref{eqresp1} from the condition in Eq. 
\eqref{eq0} has been used in \cite{fors08} for the case of nonlinear transport in a conductor.  

If we set the vector $\mathbf{l}$ to $1$ for components $a$ and $b$, and to $0$
for all other components, Eq. \eqref{eqresp1} becomes 
\begin{equation}
D^{\textrm{eq}}_{ab}= L_{ab}+L_{ba},
\label{eqfdr}
\end{equation}
which is the fluctuation dissipation relation for periodically driven systems. This equation relates fluctuations in equilibrium, as quantified by  $D^{\textrm{eq}}_{ab}$, with 
nonequilibrium response functions, as quantified by the Onsager coefficients. For the case $a=b$ we obtain $D^{\textrm{eq}}_{aa}= 2L_{aa}$ by setting the component $a$ of the vector 
$\mathbf{l}$ to $2$ and the other components to $0$  in Eq. \eqref{eqresp1}. 

\subsection{Symmetry for Onsager coefficients}

For the reciprocity relation for periodically driven systems, we also have to consider the bipartite Markov process corresponding to reversal of the protocol.
From the fluctuation theorem for currents in Eq. \eqref{eqft} we obtain
\begin{align}
\left.\frac{\partial^2 G}{\partial z_a\partial\F_b}\right|_{\mathbf{z}=\mathbf{z}^*,\mathbf{F}=\mathbf{F}^*}= &
\left.\frac{\partial^2 G^R}{\partial z_a\partial z_b}\right|_{\mathbf{z}=-\mathbf{z}^*-\mathbf{F}^*,\mathbf{F}=\mathbf{F}^*}\nonumber\\
& -\left.\frac{\partial^2 G^R}{\partial z_a\partial\F_b}\right|_{\mathbf{z}=-\mathbf{z}^*-\mathbf{F}^*,\mathbf{F}=\mathbf{F}^*}
\label{eqresp2}
\end{align}
Setting $\mathbf{z}^*=\mathbf{F}^*=0$, Eq. \eqref{eqresp2} becomes 
\begin{equation}
L_{ab}=D^{\textrm{eq}}_{ab}-L^R_{ab}.
\end{equation}
This equation together with the fluctuation dissipation relation in Eq. \eqref{eqfdr} gives the symmetry of the Onsager coefficients 
\begin{equation}
L^R_{ab}=L_{ba}.
\end{equation}
This symmetry relation is a generalization of the symmetry derived in \cite{bran15}, since our framework also accounts for the case 
of non-zero fixed thermodynamic affinities $\F_\alpha$. 

We note that this method of taking derivatives of the fluctuation theorem for
the currents to derive relations for response coefficients as in Eq. \eqref{eqresp2} has been used in \cite{andr07b} for the case of 
nonequilibrium steady states. The main difference between the derivations in this reference  and the present derivation is that for periodically 
driven systems we have to consider two scaled cumulant generating functions and, therefore, Eq. \eqref{eqresp2} alone is not 
enough to get the symmetry of Onsager coefficients, we also need Eq. \eqref{eqfdr}.

\subsection{Nonlinear coefficients}

We now show that the fluctuation theorem for the currents also implies relations between the nonlinear response coefficients. Expanding  
the current up to second order in the affinity we obtain   
\begin{equation}
J_a= \sum_bL_{ab}\F_b+\frac{1}{2}\sum_{bc}M_{a,bc}\F_b\F_c+ \textrm{O}(\F^3),
\label{eqsecondorder}
\end{equation}
where 
\begin{equation}
M_{a,bc}\equiv \left.\frac{\partial^3 G}{\partial z_a\partial\F_b\partial\F_c}\right|_{\mathbf{z}=0,\mathbf{F}=0}.
\end{equation}
The diffusion coefficient is expanded up to first order, 
\begin{equation}
D_{ab}= D_{ab}^{\textrm{eq}} +\sum_c N_{ab,c}\F_c+\textrm{O}(\F^2),
\end{equation}
where
\begin{equation}
N_{ab,c} \equiv \left.\frac{\partial^3 G}{\partial z_a\partial z_b\partial\F_c}\right|_{\mathbf{z}=0,\mathbf{F}=0}.
\end{equation}

From the  fluctuation theorem for the currents in Eq. \eqref{eqft} we see that 
the scaled cumulant generating function in equilibrium is symmetric and, hence, 
the odd cumulants associated with the currents in equilibrium are zero.   
In particular, using the fact that the third cumulant in equilibrium is zero, from Eq. \eqref{eqresp1} we obtain
\begin{equation}
M_{a,bc}+M_{b,ac}+M_{c,ab}=N_{ab,c}+N_{ac,b}+N_{bc,a}.
\label{eqresp3}
\end{equation}
Hence, the second order coefficients of the current can be expressed as first order coefficients of the diffusion 
coefficient. Furthermore, taking a further derivative with respect to $\F_c$ in Eq. \eqref{eqresp2} we obtain
\begin{equation}
M_{a,bc}+M_{a,bc}^R= N_{ab,c}+N_{ac,b}= N^R_{ab,c}+N^R_{ac,b},
\label{eqresp4}
\end{equation}
where the second equality comes from the fact that we can interchange the roles 
of original and reversed protocol in Eq. \eqref{eqresp2}.  From Eqs. \eqref{eqresp3} and 
\eqref{eqresp4} the following relation for the second order coefficient  of the current  is obtained,
\begin{equation}
M_{a,bc}+M_{b,ac}+M_{c,ab}=M^R_{a,bc}+M^R_{b,ac}+M^R_{c,ab}.
\end{equation}

In general, relation \eqref{eqresp1} shows that  higher order cumulants  at equilibrium can be expressed as
 response functions associated with lower order cumulants. Considering higher orders in Eq. 
\eqref{eqresp1} and taking further derivatives in Eq. \eqref{eqresp2} lead to relations between 
higher order response coefficients. For the case of a symmetric protocol, 
all relations for nonlinear response coefficients derived in \cite{andr07b} hold true, since their derivation relies on 
the relation $G(\mathbf{z},\mathbf{F})=G(-\mathbf{z}-\mathbf{F},\mathbf{F})$ that is valid for a symmetric protocol.

%==========================================================================
\section{Conclusion}
%==========================================================================
\label{sec5}

We have proven a fluctuation theorem for the currents for periodically driven systems. This result 
generalizes  the symmetry of the Onsager coefficients for periodically driven systems  obtained in \cite{bran15}:
our fluctuation theorem implies this symmetry, a fluctuation dissipation relation, and further relations for 
nonlinear response coefficients. This situation is akin to the previously known  fluctuation theorem for the currents for steady  
states that implies response relations.

Our results also provide a unifying framework that includes two different 
classes of periodically driven systems that have hitherto  been analyzed separately in the literature and 
that have been realized experimentally. These two classes  are small heat engines operated with periodic temperature variation and 
molecular pumps that can have fixed thermodynamic forces and, therefore, would be  out of equilibrium even without periodic driving. 

Several universal features of nonequlibrium steady states have been obtained within the framework of stochastic thermodynamics \cite{seif12}.
The fluctuation theorem for currents is one such universal feature that is now generalized to the case of periodically driven systems.  
Generalizing other results that have been established  for nonequilibrium steady states, e.g., fluctuation dissipation relations far from equilibrium \cite{seif09,baie09,alta16} and the thermodynamic uncertainty 
relation \cite{bara15a,bara16a}, to periodically driven systems constitutes an interesting direction for future work.

\begin{acknowledgments}
We thank Udo Seifert for helpful discussions and Matteo Polettini for pointing out 
a mistake in a previous version.  
\end{acknowledgments}
%==========================================================================
% References
%==========================================================================
\appendix

%==================================================================================================================================================
\section{Deterministic protocol as a limit of a stochastic protocol}
%==================================================================================================================================================
\label{appa}

In this Appendix we explain how  a continuous deterministic 
protocol can be obtained as a stochastic protocol with infinitely
many jumps. We also write down the expression of the entropy 
production for this case of a deterministic protocol. 

\subsection{Average entropy production} 

The stochastic protocol alone is a Markov process that follows the master equation
\begin{equation}
\frac{d}{dt}P^n=\gamma P^{n-1}-\gamma P^n,
\end{equation}
where $P^n=\sum_iP_i^n$ and we set $\gamma^n=\gamma$ for $n=0,1,\ldots,N-1$. If we consider a random variable $X_{\textrm{ext}}$  that counts the 
number of jumps of the external protocol, a standard calculation gives 
\begin{equation}
v_{\textrm{ext}}\equiv\langle X_{\textrm{ext}}\rangle/t= \gamma/N
\end{equation}
and 
\begin{equation}
D_{\textrm{ext}}\equiv\langle \left(X_{\textrm{ext}}-\langle X_{\textrm{ext}}\rangle\right)^2\rangle/t= \gamma/N^2.
\end{equation}
By setting $\gamma=N/\tau$ and taking the limit $N\to \infty$, the stochastic protocol becomes deterministic
with a speed $v_{\textrm{ext}}=\tau^{-1}$ and a dispersion $D_{\textrm{ext}}= (\tau N)^{-1}\to 0$. In this limit 
the transition rates $w_{ij}^n$ become $w_{ij}(t)$, where $t= n \tau/N$. The periodicity  condition $w_i^n=w_i^{n+N}$
changes to $w_{ij}(t)=w_{ij}(t+\tau)$.

The master equation in this limit then becomes 
\begin{equation}
\frac{d}{dt}R_i(t)= \sum_j\left[ R_j(t)w_{ji}(t)-R_i(t)w_{ij}(t)\right],
\label{eqmeq2}
\end{equation}
where $R_i(t)$ is the probability to be in state $i$ at time $t$. The generalized detailed balance relation in Eq. \eqref{eqgendb} changes to 
\begin{equation}
\ln\frac{w_{ij}(t)}{w_{ji}(t)}= \beta(t)\left[E_i(t)-E_j(t)+(\beta_c)^{-1}\F_\alpha d_{ij}^{(\alpha)}\right],
\label{eqgendb2}
\end{equation}
where $\beta(t)= \beta_c[1-\F_q h(t)]$ and $E_i(t)= E_i+ \Delta E  f_i(t)$. Comparing with the stochastic protocol, the 
functions $h(t)$ and $E_i(t)$ fulfill the relations $h(t=\tau n/N)= h^n$ and $E_i(t=\tau n/N)= E_i^n$, where $E_i^n$ is given in Eq. \eqref{eqE} and $h^n$ is given in Eq. \eqref{eqbeta}  

In the long time limit the system reaches a periodic steady state characterized by the probability  $R^*_i(t)=R_i^*(t+\tau)$.
For the comparison of this probability with the stationary probability of the bipartite process $P_i^n$, we define the conditional stationary 
probability of the system being in state $i$ given the protocol is in state $n$ $P(i|n)\equiv P_i^n/P^n$, where 
the stationary probability of the protocol is $P^n=1/N$. It can be shown that the conditional probability 
of the bipartite Markov process $P(i|n)$ tends to $R^*_i(t=n \tau/N)$ in the limit $N\to\infty$ \cite{bara16a}. 

The elementary current $X^*_{ij}$, analogous to $\sum_nX^n_{ij}$ for a stochastic protocol, is a random variable that increases by one if a jump from $i$ to $j$ takes place and that decreases by one if 
a jump from $j$ to $i$ takes place. The average current  
\begin{equation}
J^*_{ij}\equiv\lim_{t\to\infty}\frac{\langle X^*_{ij}\rangle}{t},
\end{equation}
is given by
\begin{equation}
J^*_{ij}= \frac{1}{\tau}\int_0^\tau\left[R^{*}_i(t)w_{ij}(t)-R^{*}_j(t)w_{ji}(t)\right]dt\equiv  \frac{1}{\tau}\int_0^\tau J_{ij}(t)dt.
\end{equation}
Using Eq. \eqref{eqelem}, this expression can be compared to the following expression for the stochastic protocol 
\begin{equation}
J_{ij}\equiv \sum_n J_{ij}^n=\frac{1}{N}\sum_n\left[P(i|n)w_{ij}^n-P(j|n)w_{ji}^n\right].
\end{equation}
Comparing with $J^*_{ij}$, we obtain that the convergence $P(i|n)\to R^{*}_i(t=n\tau/N)$ in the limit $N\to \infty$, implies $J_{ij}\to J^*_{ij}$. 

The entropy production in Eq. \eqref{eqent} changes to 
\begin{align}
\sigma^* & \equiv \frac{1}{\tau}\int_0^\tau\sum_{ij} R^{*}_i(t)w_{ij}(t)\ln\frac{w_{ij}(t)}{w_{ji}(t)}\nonumber\\
&=\F_q J^*_q+ \F_eJ^*_e+\sum_\alpha \F_\alpha J^*_\alpha\ge 0 .
\end{align}
where 
\begin{equation}
J^*_\alpha\equiv \frac{1}{\tau}\int_0^\tau \sum_{i<j} (\beta_c)^{-1}\beta(t)J_{ij}(t)d_{ij}^{(\alpha)}dt,
\label{eqjalpha2}
\end{equation}
\begin{equation}
J^*_q\equiv \frac{1}{\tau}\int_0^\tau\sum_{i<j} J_{ij}(t) h(t)\beta_c[E_j(t)-E_i(t)]dt,
\label{eqjq2} 
\end{equation}
and 
\begin{equation}
J^*_e \equiv  \frac{1}{\tau}\int_0^\tau\sum_{i<j} J_{ij}(t) [f_i(t)-f_j(t)]dt.
\label{eqje2} 
\end{equation}

The fact that the stationary probability of the bipartite process converges to $R_i^*(t)$ suggests that such 
convergence should also take place for current fluctuations, as characterized by the scaled cumulant generating function.
Furthermore, an expression for the large deviation function characterizing fluctuations of currents in periodically driven systems 
with a deterministic protocol in terms of $R_i^*(t)$ has been recently proposed in \cite{rost17}.  Such an expression provides further 
evidence for this convergence for current fluctuations. We now illustrate the convergence of the scaled cumulant generating 
function for a specific model analyzed in \cite{bara16a}.

\subsection{Current fluctuations}

\begin{figure}
\includegraphics[width=75mm]{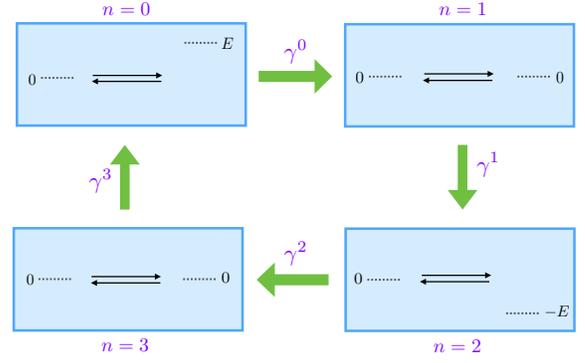}
\vspace{-2mm}
\caption{(Color online) Model with symmetric protocol for $N=4$.
}
\label{fig3} 
\end{figure}

The model with a symmetric protocol illustrated in Fig. \ref{fig3} is defined as follows. The system has two states, one with energy zero 
and the other with energy $E^n=\F_e\cos(2\pi n/N)$. The temperature is constant and set to $\beta^n=1$. The transition 
rates of the protocol are $\gamma^n$. For the comparison with a deterministic protocol we set $\gamma^n=\gamma$.
The transition rate from the state with energy $0$ to the state with energy $E^n$ is set to $k\textrm{e}^{-E^n/2}$, 
while the reversed transition rate is $k\textrm{e}^{E^n/2}$. The scaled cumulant generating function $G_s(z)$
can be obtained by calculating the eigenvalue of the modified generator from Sec. \ref{sec3}. 

We now consider the deterministic version of the model. The probability vector $\mathbf{R}(t)$ has two components, 
with the first component as the probability that the system in the state with energy $0$ and the second as the probability 
that the system is in the state with energy  $E(t)= \F_e \cos(t)$. The transition rate from the state with energy $0$ to the 
states with energy $E(t)$ is $k\textrm{e}^{-E(t)/2}$, whereas the reversed transition rate is $k\textrm{e}^{E(t)/2}$.

\begin{figure}
\includegraphics[width=75mm]{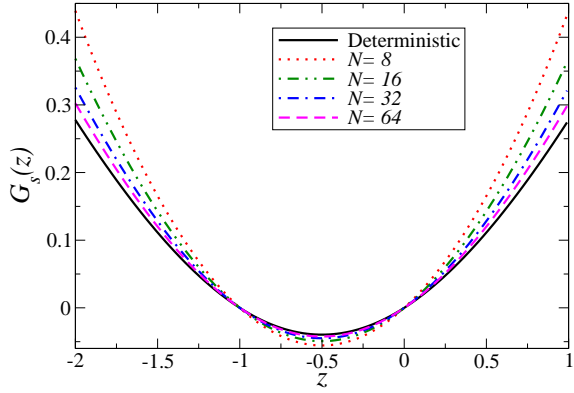}
\vspace{-2mm}
\caption{(Color online) Scaled cumulant generating functions for a deterministic protocol and for a stochastic protocol. The scaled 
cumulant generating function for a stochastic protocol tends to the scaled cumulant generating function for the deterministic protocol 
with increasing $N$. The parameters of the model with a symmetric protocol are set to $\gamma=2\pi/N$, $k=1$ and $\F_e=2$, where the parameter 
$\gamma$ is valid only for the stochastic protocol.
}
\label{figlast} 
\end{figure}

The probability vector $\mathbf{R}(X_s^*,t)$ gives the 
probabilities that the system is a certain state with the entropy current given by $X_s^*$. Defining the Laplace 
transform  $\mathbf{R}(z,t)=\sum_{X_s^*}\mathbf{R}(X_s^*,t)\textrm{e}^{zX_s^*}$ and using the master equation \eqref{eqmeq2}, we obtain 
\begin{equation}
\frac{d}{dt}\mathbf{R}(z,t)= \mathcal{L}(z,t)\mathbf{R}(z,t),
\end{equation}
where 
\begin{equation}
\mathcal{L}(\mathbf{z},t)=\left(\begin{array}{cc}
-k\textrm{e}^{E(t)/2} & k\textrm{e}^{-E(t)/2} \textrm{e}^{-z E(t)} \\
k\textrm{e}^{E(t)/2} \textrm{e}^{z E(t)} & -k\textrm{e}^{-E(t)/2} \\
\end{array}\right).
\end{equation}
The scaled cumulant generating function is given by
\begin{align}
&& G_s^*(z)  \equiv \lim_{t\to\infty}\frac{1}{t}\ln\langle\exp(zX_s^*)\rangle\nonumber\\
&& =  \lim_{t\to\infty}\frac{1}{t}\ln[R_1(z,t)+R_2(z,t)]. 
\end{align} 
where $R_i(z,t)$ is the component of the vector $\mathbf{R}(z,t)$. Using Floquet theory \cite{klau08}, the scaled cumulant 
generating function $G_s^*(z)$ is given by the maximal Floquet exponent associated with $\mathcal{L}(\mathbf{z},t)$.
We have calculated this maximal Floquet exponent following the numerical method explained in \cite{klau08}. In Fig. \ref{figlast}, we show the convergence
of the scaled cumulant generating function obtained with the stochastic protocol with increasing $N$ to $G_s^*(z)$. 
We note that, to our knowledge, a rigorous proof of the large deviation principle for arbitrary currents in periodically driven systems with 
deterministic protocols is still lacking. However, it is reasonable to expect that beyond the example analyzed here this scaled cumulant generating function 
is given by a maximal Floquet exponent.    

%==================================================================================================================================================
\section{Equality between characteristic polynomials}
%==================================================================================================================================================
\label{appb}

The scaled cumulant generating function $G(\mathbf{z})$ is a root of the characteristic polynomial associated with $\mathcal{L}(\mathbf{z})$. This polynomial is given by the 
determinant of the matrix $\mathcal{L}(\mathbf{z})-\mathcal{I}x$, where $\mathcal{I}$ is the identity matrix with dimension $\Omega\times N$ and $x$ is the variable of the 
polynomial. From Eq. \eqref{eqgensto1} this matrix   takes the form 
\begin{equation}
\left(\begin{array}{ccccc}
\mathcal{\mathcal{L}}_{0}(\mathbf{z})-\mathbf{D}_0& 0 & \ldots & \mathbf{\Gamma}_{N-1}\\
\mathbf{\Gamma}_0 & \mathcal{\mathcal{L}}_{1}(\mathbf{z})-\mathbf{D}_1 &  \ldots  & 0\\
0 & \mathbf{\Gamma}_1 &  \ldots  & 0\\
\vdots & \vdots & \ddots & \vdots \\
0 & 0 & \ldots &  \mathcal{\mathcal{L}}_{N-1}(\mathbf{z})-\mathbf{D}_{N-1}
\end{array}\right).
\label{eqgenstoapp1}
\end{equation}
where $\mathbf{D}_n=\mathbf{I}(\gamma^n+x)$. Furthermore, from Eqs. \eqref{eqgensto3}
and \eqref{eqsmallsym}, the transpose of the matrix $\mathcal{D}\mathcal{L}^R(-\mathbf{F}-\mathbf{z})\mathcal{D}^{-1}-\mathcal{I}x$, where 
$\mathcal{D}$ is the diagonal matrix from Eq. \eqref{eqsimi}, reads 
\begin{equation}
\left(\begin{array}{ccccc}
\mathcal{\mathcal{L}}_{0}(\mathbf{z})-\mathbf{D}_0& 0 & \ldots & \mathbf{\Gamma}_{0}\\
\mathbf{\Gamma}_1 & \mathcal{\mathcal{L}}_{1}(\mathbf{z})-\mathbf{D}_1 &  \ldots  & 0\\
0 & \mathbf{\Gamma}_2 &  \ldots  & 0\\
\vdots & \vdots & \ddots & \vdots \\
0 & 0 & \ldots &  \mathcal{\mathcal{L}}_{N-1}(\mathbf{z})-\mathbf{D}_{N-1}
\end{array}\right).
\label{eqgenstoapp2}
\end{equation}
In order to show that the matrices in Eqs. \eqref{eqgenstoapp1} and \eqref{eqgenstoapp2} have the same determinant we consider Leibniz formula for determinants (Eq. 0.3.2.1 in \cite{matrixbook}), 
where the determinant is written as a sum over all $(\Omega\times N)!$ permutations of the elements. In a graphical representation of these terms, where the 
states of the bipartite process $(i,n)$ are vertices and nonzero transition rates are edges, there are diagonal terms and cyclic permutations with sizes that range from 2 up to  
$\Omega\times N$ (see \cite{andr07c,bara12a}). For the case of the matrices in Eqs. \eqref{eqgenstoapp1} and \eqref{eqgenstoapp2} there 
are two kinds of cycles. First there are cycles that do not contain external jumps that lead to a change in the external protocol. In this case, since the diagonal blocks 
in Eqs. \eqref{eqgenstoapp1} and \eqref{eqgenstoapp2} are identical, the contribution to the determinants coming from these cycles must be the same for both matrices.
Second, there are cycles that contain external jumps. In this case, since the external jumps are irreversible, all such cycles must go through  all external states in order 
to close the cycle. For both matrices, the contribution to these cycles due to the external jumps is the same and given by $\prod_{n=0}^{N-1}\gamma^n$. We thus conclude 
that all terms contributing to the determinant, namely, diagonal terms, cycles containing only internal jumps and cycles containing external jumps, are exactly the same 
for the matrices in Eqs. \eqref{eqgenstoapp1} and \eqref{eqgenstoapp2}. Hence, the determinants of these matrices are identical, which leads to the symmetry 
$G(\mathbf{z})=G^R(-\mathbf{F}-\mathbf{z})$.

\end{document}